\documentclass[aps,twocolumn,showpacs,showkeys]{revtex4}
\usepackage[pdftex]{graphicx}
\usepackage{amssymb}
\usepackage{txfonts}
\def\ZMG{Zn$_{1-x}$Mg$_x$O}

\begin{document}
\title {Microstrain and defects in polycrystalline \ZMG\/ 
(0 $\leqslant x \leqslant$ 0.15) studied by X-ray diffraction, 
and optical and Raman spectroscopies}

\author{Young-Il Kim and Ram Seshadri}
\affiliation{Materials Department and Materials Research Laboratory, University of 
California, Santa Barbara, California 93106, USA}
\date{Received \today}

\begin{abstract}

The properties of polycrystalline wurtzite \ZMG\/ (0\,$\leqslant$\,$x$\,$\leqslant$\,0.15) have been studied, 
by powder X-ray diffraction, including an analysis of the 
X-ray line-broadening, and by luminescence, absorption,
and Raman spectroscopies. 
We have previously established from synchrotron X-ray difffraction 
that with increasing Mg-concentration, the $c$-axis compresses,   
and the off-center cation displacement within 
each tetrahedral (Zn,Mg)O$_4$ unit decreases. 
Here we perform a size-strain line-broadening analysis 
of the XRD peaks, which reveals that the Mg-substitution 
reduces the coherent crystallite size and also develops 
the lattice microstrain. 
The optical properties of the samples have been characterized by 
diffuse-reflectance spectroscopy and fluorimetry. 
Both the optical band gap and the band-edge emission energies 
gradually increase with the Mg-concentration in \ZMG\/.
The Mg-substituted samples show broader band tails near the absorption edge, 
due to the increase of crystal imperfections. 
The peak broadening of the $E_{2}^\mathrm{high}$ Raman mode, upon the Mg-substitution, is also ascribed to the phonon shortening lifetime 
mechanism \textit{via} crystal defects. 

\end{abstract}



\maketitle

\section{INTRODUCTION}

Zinc oxide (ZnO) is well known for the diverse uses deriving 
from semiconducting, piezoelectric, pyroelectric, photoluminescent, 
and photocatalytic characteristics.\cite{Ozgur,Pearton,Look} 
Among other properties, the polar crystal lattice of ZnO provides 
great potential for developing two-dimensional electron gas (2DEG) 
\textit{via} the construction of heterojunction structures.  
A 2DEG can be created at heterojunction interface that bears 
a polarization gradient, with both high carrier concentration 
and high carrier mobility.\cite{Davies}
In the recent experiments on II-VI or III-V semiconductors, 
2DEG have been well utilized for exploring novel transistor 
devices\cite{Tampo,Rajan} as well as the quantum Hall effect\cite{Tsukazaki}.

The alloy \ZMG\/ is a suitable system for exploiting the ZnO-based 2DEG, 
since the Mg-substitution provides an efficient way to tune the 
polarization strength of ZnO crystal with only minimal changes 
of lattice dimensions.\cite{Kim1} The polarization in \ZMG\/ and the 
2DEG quality of ZnO/\ZMG\/ heterostructure will depend mostly on 
the crystal structure details of \ZMG\/ alloy. Hence the sensible 
design of ZnO/\ZMG\/ heterostructure devices requires a comprehesive 
knowledge of structural evolutions across \ZMG\/ solid solutions.

The composition-dependent structure evolution of \ZMG\/ alloy using 
synchrotron X-ray diffraction has been presented by us 
previously.\cite{Kim1,Kim2} The use of high-flux and high-energy 
synchrotron radiation enabled us to trace small changes of 
wurtzite parameters ($a$, $c$, and $u$) in \ZMG\/ solid solutions. 
In addition line-broadening analysis\cite{Scardi,Langford,Balzar} 
of powder XRD patterns provides information on the crystallite 
morphology and Mg patterns in \ZMG\/ phases. 
The XRD analyses show that the Mg substitution modifies the 
macroscopic distortion of the hexagonal lattice through the 
enhanced bond ionicity, and also that the static polarization 
in the crystal can be gradually varied, in parallel with the 
internal tetrahedral distortion, as a function of Mg content. 
The effects of Mg-substitution on the optical properties and 
Raman spectra are also discussed here. 
Raman spectroscopy, which is known to be a powerful probe of 
compositional disorder and/or the strain within semiconductor 
alloys,\cite{Richter,Tiong} corroborates the findings from XRD analyses.

\section{EXPERIMENTS}

Polycrystalline \ZMG\/ ($x$ = 0, 0.05, 0.10, 0.15 and 0.20) samples were 
prepared by an oxalate co-precipitation way.\cite{Kim1,Kim2} 
Zinc acetate and magnesium acetate were dissolved together in deionized 
water in the stoichiometric cation ratios, and mixed with separately 
prepared oxalic acid solution. The resulting precipitates were thoroughly 
washed and dried at 60$^{\circ}$C for 4\,h to produce white and 
crystalline Zn$_{1-x}$Mg$_{x}$(C$_2$O$_4$)$\cdot$2H$_2$O powders. 
The oxalate dihydrates were converted to the oxides \ZMG\/ by heating in air 
at 550$^{\circ}$C for 24\,h. Powder X-ray diffraction (XRD) identified the 
phase pure Zn$_{1-x}$Mg$_{x}$(C$_2$O$_4$)$\cdot$2H$_2$O and \ZMG\/ after 
the heating steps of 60$^{\circ}$C and 550$^{\circ}$C, respectively. 
Thermogravimetry of Zn$_{1-x}$Mg$_{x}$(C$_2$O$_4$)$\cdot$2H$_2$O, 
in air up to 1000$^{\circ}$C, confirmed that the oxide \ZMG\/ phases 
are formed through well-defined processes of dehydration and the oxalate 
decomposition. 

For the crystal structure determinations by Rietveld analysis, 
synchrotron XRD patterns of \ZMG\/ were measured on beamline 11-ID-B 
of the Advanced Photon Source (Argonne National Laboratory, USA) with an 
X-ray wavelength of 0.13648\,\AA. The XRD patterns for the 
line-broadening analysis were obtained using an in-house X-ray diffractometer 
(Philips X'PERT MPD, Cu $K\alphaup_{1,2}$). 
Instrumental broadening was corrected using the data measured 
for LaB$_6$ (grain size $\approx$10\,$\muup$m). 
Diffuse-reflectance absorption spectra were recorded for \ZMG\/ 
in the wavelength range of 220$-$800\,nm using a Shimadzu UV-3600 
spectrophotometer equipped with an ISR-3100 integrating sphere. 
The optical band gap was determined by extrapolating the linear part of 
the absorption edge to zero-absorption level. 
Photoluminescence of \ZMG\/ powders were studied using 
a Perkin-Elmer LS55 luminescence spectrometer at room temperature. 
Emission spectra were recorded employing an excitation wavelength of 340\,nm. 
Raman spectra at room temperature were acquired using an optical 
microprobe fitted with a single monochromator (Jobin-Yvon, T64000). 
An Ar$^+$ laser ($\lambdaup$ = 488\,nm) excitation was used with a beam power 
of 50\,mW and a spot size of $\approx$2\,$\muup$m. 
The $E_2^\mathrm{high}$ phonon mode was used for detailed peak profile analyses. 
The background was removed following Shirley\cite{Shirley}, and 
Breit-Wigner-Fano type peak fitting\cite{Yoshikawa} was carried out to determine 
peak position and the width.

\section{RESULTS AND DISCUSSION}

\begin{figure}
\smallskip \centering \includegraphics[width=7.5cm]{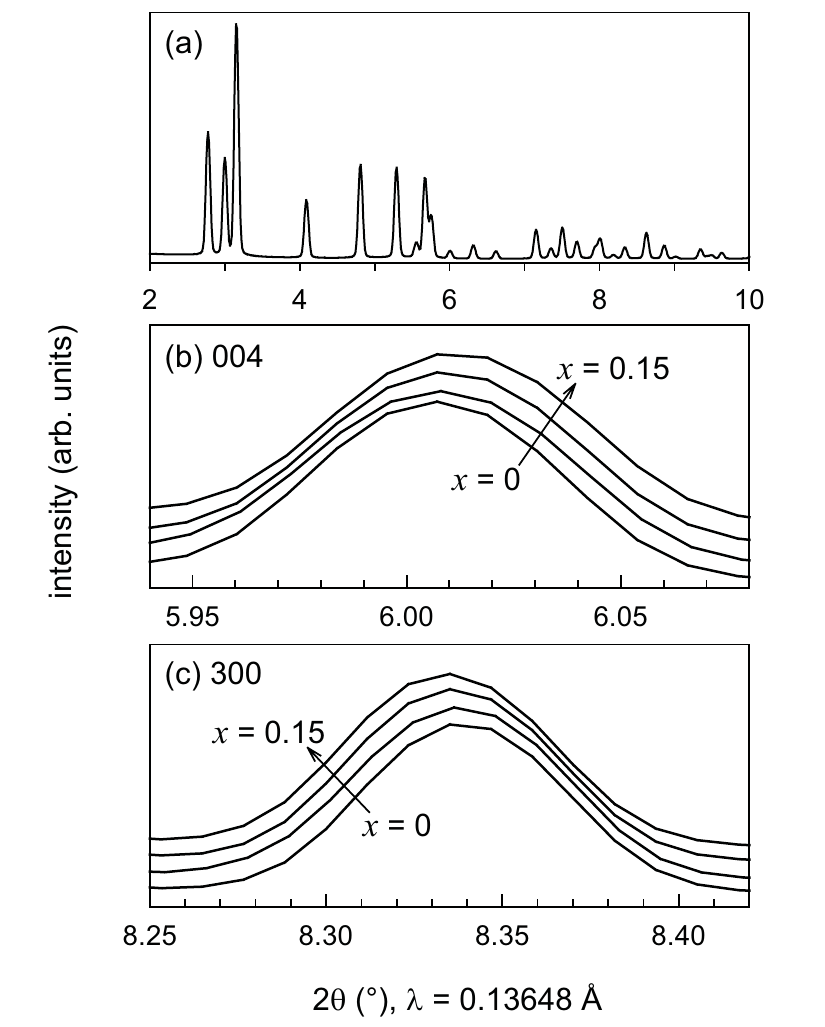}
\caption{(a) Synchrotron XRD pattern for ZnO and the zoomed views 
of (b) in-plane 004 and (c) out-of-plane 300 XRD peaks for 
\ZMG\/ ($x$ = 0, 0.05, 0.10, and 0.15).}
\label{fig:xrd}
\end{figure}

The synchrotron XRD pattern for ZnO is shown in Figure\,\ref{fig:xrd}a, 
which is consistent with the wurtzite-type hexagonal structure. 
All the Mg-substituted phases ($x$ = 0.05, 0.10, 0.15, and 0.20) presented nearly
indistinguishable XRD patterns to that of ZnO, but the $x$ = 0.20 phase was found 
to contain cubic MgO as a secondary phase. Therefore the solubility of Mg in \ZMG\/ is estimated to be slightly higher than 15\,\%, under the preparational conditions employed here. 
The ionic radii of 4-coordinate Mg$^{2+}$(0.57\,\AA) and Zn$^{2+}$(0.60\,\AA) 
are similar\cite{Shannon} but the two cations will demand distinct extents of 
geometric distortions in the tetrahedral environment. 
The resulting structural evolution in \ZMG\/ solid solutions can be first  
recognized from the peak positions of 00$l$ and $h$00 reflections. 
As displayed in Figure\,\ref{fig:xrd}b and c, the 004 and 300 diffraction peaks 
of \ZMG\/ are shifted to higher and lower angles, respectively, as $x$ increases. 
Namely, the hexagonal lattice constant $a$ increases and $c$ decreases with the 
progress of Mg-substitution in ZnO. 

For the quantitative determinations of structural parameters, 
Rietveld refinements were performed for \ZMG\/ using a structure 
model based on wurtzite ZnO; space group $P6_3mc$, Zn/Mg at 
($\frac{1}{3}$\,$\frac{2}{3}$\,0) and O at ($\frac{1}{3}$\,$\frac{2}{3}$\,$u$).\cite{Kim1} 
Rietveld refinements confirmed that the wurtzite structure is retained for 
all the \ZMG\/ phases studied here, and also that the 
Mg-substitution in ZnO accompanies the expansion of $ab$-dimension, 
compression of $c$-dimension, and consequently the decrease of $c/a$ ratio. 
The $a$ and $c$ parameters follow approximately, the relationships: 
$a(\mathrm{\AA})=3.2503+0.0118x$ and $c(\mathrm{\AA})=5.2072-0.0232x$, 
as functions of $x$.
The variations of $a$, $c$, and $c/a$ in \ZMG\/ solid solutions are 
consistent with the general trend found in the binary wurtzite 
family, that is, $c/a$ ratio becomes smaller as the bonding character 
becomes more ionic.\cite{Schulz} 
It is the increased ionicity that Mg$^{2+}$ provides in \ZMG\/ 
which results in the $c$-axis compression of the hexagonal lattice.

The internal tetrahedral distortion and the spontaneous polarization 
in \ZMG\/ can be assessed using the atomic position parameter $u$. 
The four nearest cation-anion pairs are equidistant 
if $u=\frac{1}{3}(\frac{a}{c})^2+\frac{1}{4}$, 
whereas the charge separation in each tetrahedral unit will vanish if 
$u=\frac{3}{8}$. 
The Rietveld-refined $u$ monotonously decreased from 0.3829 for ZnO to 
0.3819 for Zn$_{0.85}$Mg$_{0.15}$O.\cite{Kim1} This indicates that the tetrahedral 
distortion in ZnO is gradually relieved by alloying with MgO, and that ionic 
polarization of \ZMG\/ should decrease with the Mg concentration. 
This was also supported by pair-distribution-function studies of \ZMG.\cite{Kim2} 

\begin{figure}
\smallskip \centering \includegraphics[width=7.5cm]{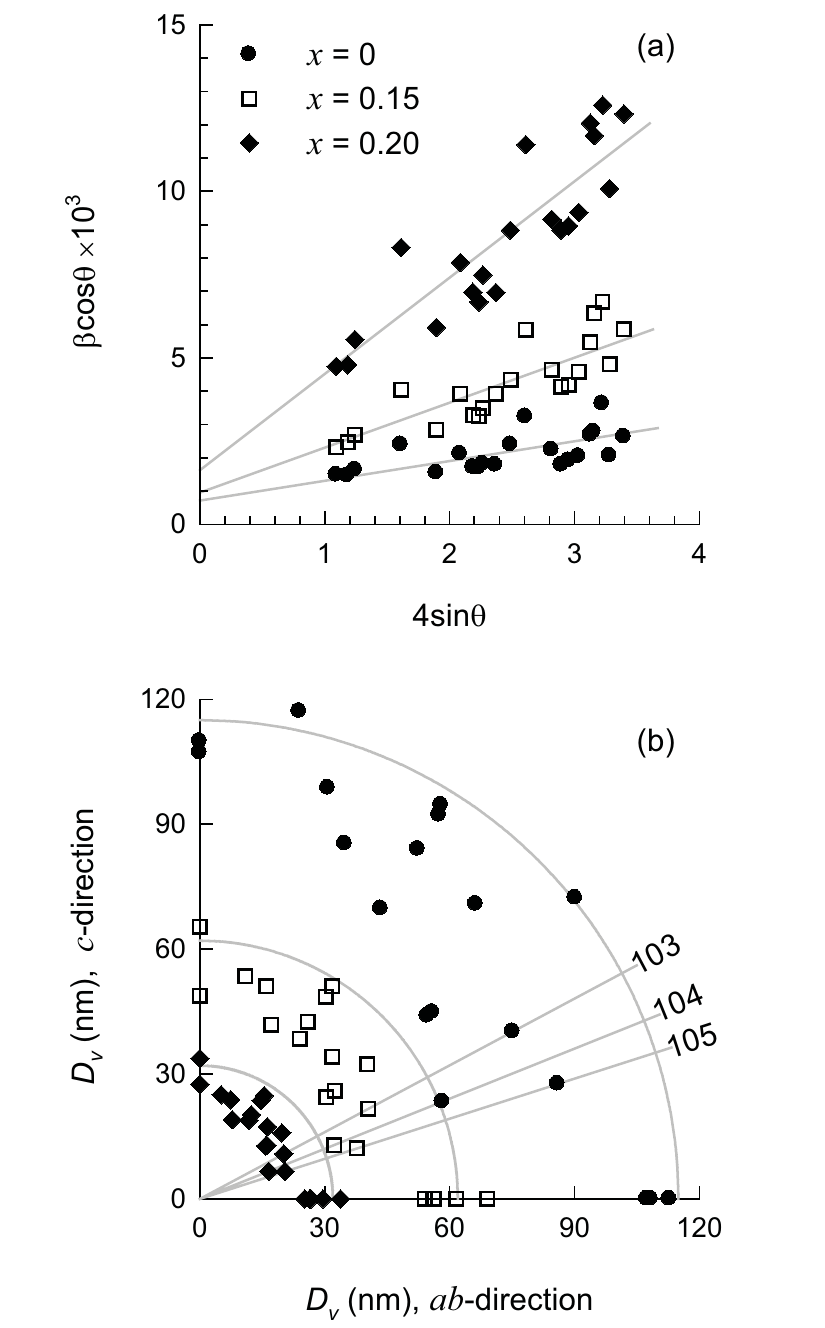}
\caption{(a) Williamson-Hall plots and (b) polar diagrams of apparent 
crystallite size $D_v$ for \ZMG\/ with $x$ = 0, 0.15, and 0.20.}
\label{fig:WH}
\end{figure}

\begin{table}
\caption{Crystallite size and strain of \ZMG\/ powders as 
determined by the size-strain line-broadening analysis 
using the simplified integral-breadth method.}
\begin{ruledtabular}
\begin{tabular}{llll}
$x$  & $D_v$(nm) & $D_a$(nm) & $\epsilonup$(\%) \\
\hline
0    & 96(3)     & 69(3)     & $<$0.01       \\
0.05 & 76(3)     & 57(3)     & 0.020(6)       \\
0.10 & 70(2)     & 51(3)     & 0.028(5)       \\
0.15 & 62(2)     & 47(2)     & 0.041(4)       \\
0.20 & 33(1)     & 27(1)     & 0.099(7)       \\
\end{tabular}
\end{ruledtabular}
$D_v$: volume-weighted crystallite size.\\
$D_a$: surface area-weighted crystallite size.\\
$\epsilonup$: strain averaged over the distance $D_v/2$.
\label{tab:wh}
\end{table}

In order to examine the effect of Mg-substitution on the microstructure and crystallite morphology, XRD line-broadening analysis was performed 
for \ZMG\/ ($x$ = 0, 0.05, 0.10, 0.15 and 0.20). 
As noted ahead the sample with $x=0.20$ showed an indication of MgO segregation, which implies that the remaining wurtzite phase is nearly saturated with Mg. The line-broadening data for $x=0.20$ 
can therefore be taken as the morphological characteristics at the solubility limit. 
The integral-breadth ($\betaup$) of each peak was calculated using 
the full-width-at-half-maximum and the shape factor $m$ of 
split-Pearson VII function, with an aid of the software \textsc{Breadth}\cite{Balzar}.

Figure\,\ref{fig:WH} shows the Williamson-Hall (W-H) plots\cite{Williamson} for the 
three samples of $x$ = 0, 0.15, and 0.20, together with the polar diagrams of crystallite 
size. The W-H plot is based on the following function,
\begin{equation} 
\betaup_i\cos\thetaup_i=\frac{\lambdaup}{D_v}+4\epsilonup\sin\thetaup_i
\end{equation}  
\noindent where $\betaup_i$ is the integral-breadth (in radians 2{$\thetaup$}) 
of the $i$th Bragg reflection positioned at 2$\thetaup_i$. 
The slope and the ordinate intercept of the W-H plot can be used to 
quantify crystallite size ($D_v$) and strain($\epsilonup$) contributions 
to peak broadening. 

In Figure\,\ref{fig:WH}a, the data points on each W-H plot are 
more or less scattered around a mean straight line. A close investigation 
of ZnO data revealed that the 00$l$ or $h$00 type peak groups have 
much narrower widths than the $h$0$l$ ones. 
In fact, the uniaxial (00$l$, $h$00) and the off-axial ($h$0$l$) diffraction 
peak groups fall on their own straight lines. 
They have different slopes but nearly the same intercept, implying that 
the ZnO crystallites are reasonably isotropic, but have considerable 
amount of stacking faults. The polar diagrams in Figure\,\ref{fig:WH}b displays 
the crystallite dimensions along different crystallographic directions. 
In common for the three samples, $D_v$ is largest along the $h$00, $hk$0 or 00$l$-directions, and becomes smaller in the 
$hkl$ and $h$0$l$ directions. Particularly, the 10$l$ peaks showed pronounced 
peak broadening. However, there is found no signficant anisotropic line-broadening  
caused by the Mg-substitution, indicating that the Mg is distributed in \ZMG\/ in 
a random manner.

Table\,\ref{tab:wh} lists the crystallite sizes and microstrain evaluated 
by using \textsc{Breadth}. Mg-substitution in ZnO decreases the 
XRD-coherent crystallite size and at the same time introduces greater microstrain. 
The Zn/Mg exchange itself will affect the periodicity of the 
atomic arrangement and also induces various types of structural defects and 
microstrain as well. 
This will be important in the following consideration of optical
properties.

\begin{figure}
\smallskip \centering \includegraphics[width=7.5cm]{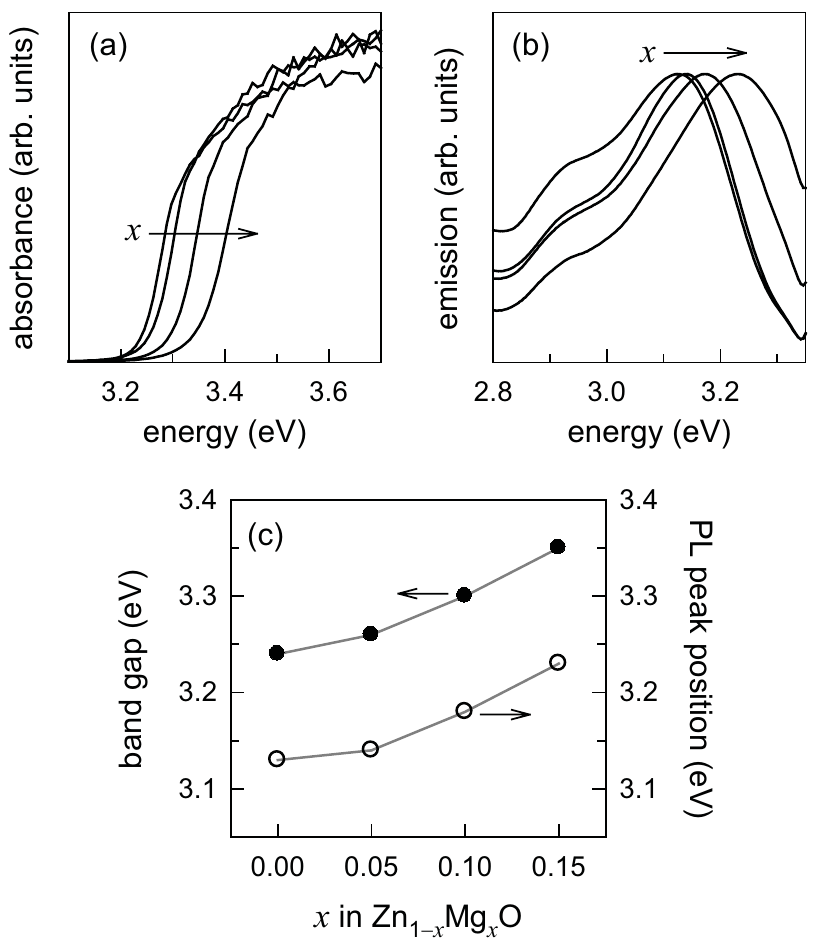}
\caption{Optical properties of polycrystalline \ZMG\/ 
($x$ = 0, 0.05, 0.10 and 0.15) 
measured at 20$^\circ$C, (a) diffuse-reflection absorption, 
(b) photoluminescence excited at 340 nm (3.65 eV), and 
(c) band gap and photoluminescence peak energies 
as functions of Mg content.} 
\label{fig:opt}
\end{figure}

Figure\,\ref{fig:opt} shows the absorption and photoluminescence spectra of 
\ZMG\/ powders, measured at room temperature. Clearly, both the band gap ($E_g$) 
transition (Figure\,\ref{fig:opt}a) and the band-edge emission 
(Figure\,\ref{fig:opt}b) are blue-shifted as the Mg content increases 
in the \ZMG\/. 
Although the solid solution of MgO$-$ZnO only involves isovalent substitution, 
the electronic structure of ZnO is substantially modified by 
the incorporation of Mg. The conduction band of wurtzite ZnO is 
mainly contributed from the Zn 4$s$ orbital. However in the \ZMG\/, 
the Mg 3$s$ orbitals contribute as well.
Since the Mg 3$s$ orbital lies at a higher energy 
level than the Zn 4$s$, the Mg-substitution in ZnO will move the 
conduction band-edge upward, thereby widening the band gap.

\begin{figure}
\includegraphics[width=7.5cm]{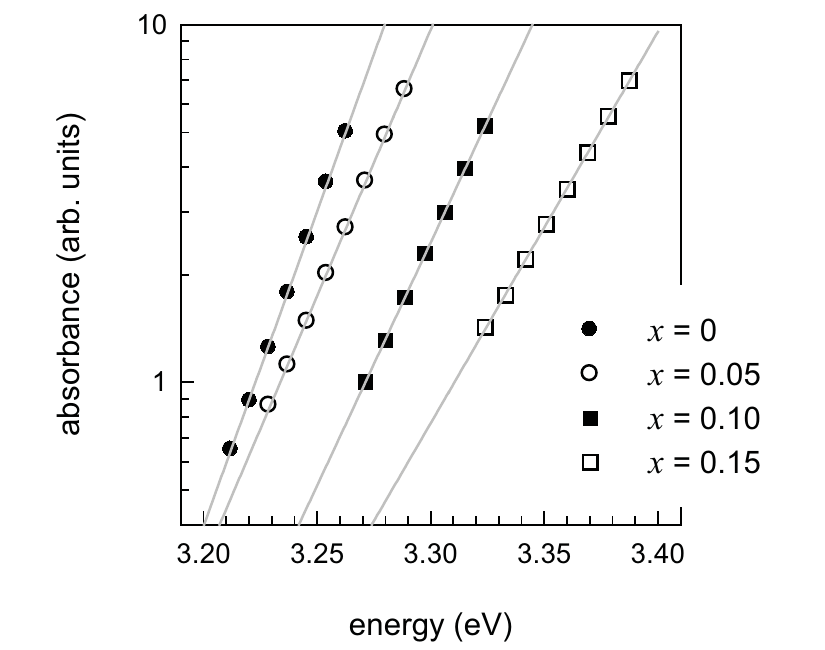}
\caption{Semilogarithmic plots Log $A$ vs. $h\nuup$ near the 
band-edge for \ZMG\/, $x$ = 0, 0.05, 0.10, and 0.15.}
\label{fig:abs}
\end{figure}

In order to evaluate the near band-edge characteristics of the 
\ZMG\/ powders, we used a description by  Pankove\cite{Pankove} 
of the absorbance ($A$) in the near band-edge region:
\begin{equation} 
A \propto \exp\frac{h\nuup}{E_0}
\end{equation}  
where $E_0$ is an empirical parameter having dimensions of energy and 
describing the width of the localized states in the band gap but not 
their energy position.
It can be regarded that $E_0$ accounts for the effects of all possible 
defects (point, line, and planar dislocations).
In Figure\,\ref{fig:abs} are shown the absorbance of \ZMG\/ powders 
as functions of the incident photon energy in the near-edge region. 
For all the compositions the absorbance shows the expected exponential 
dependence in the near band-edge regime. It is also observed that the 
value $E_0$ increases with $x$. As determined from the slope of linear fit, 
the $E_0$ widths are 24.7 ($x=0$), 29.2 ($x=0.05$), 31.9 ($x=0.10$), 
and 39.7\,meV ($x=0.15$). Such a change in band tail points out the 
role of Mg as the defect centers in the \ZMG\/ lattice. 
Not surprisingly the long range crystallinity of ZnO lattice will be 
disturbed in the course of cation substitution, by the creation of point 
defects, dislocations, impurities, and grain boundaries. 

\begin{figure}
\smallskip \centering \includegraphics[width=7.5cm]{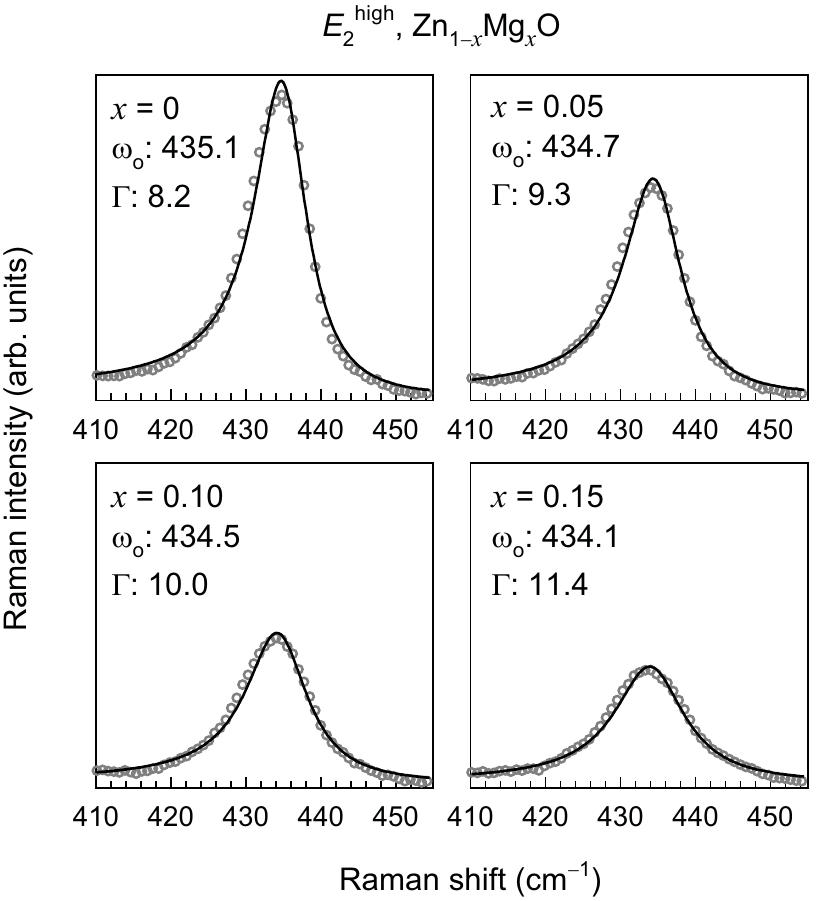}
\caption{Profile shape fittings of $E_2^\mathrm{high}$ Raman peaks for \ZMG\/ 
using the Breit-Wigner-Fano function (experimental: circles, fit: line, 
fit parameters are in cm$^{-1}$). Same intensity scale is used for all the four panels.} 
\label{fig:Raman}
\end{figure}
The $E_2^\mathrm{high}$ Raman mode of \ZMG\/ powders were 
analyzed by peak profile analyses.
In all cases of $x$ = 0$-$0.15, the $E_2^\mathrm{high}$ peaks of \ZMG\/ 
are best represented by the Breit-Wigner-Fano (BWF) lineshape\cite{Yoshikawa}, 
which is of Lorentzian type, but having the asymmetry character.
As shown in Figure\,\ref{fig:Raman}, each Raman line can be successfully 
fit to a sinlge BWF component, using the parameters of 
energy position ($\omegaup_0$), full-width-at-half-maximum ($\Gamma$), 
BWF coupling coefficient (asymmetry term), and the maximum 
peak intensity. 

As given on Figure\,\ref{fig:Raman}, the $\omegaup_0$ and $\Gamma$ of 
the \ZMG\/ $E_2^\mathrm{high}$ mode systematically change with the Mg-concnetration. 
The redshift of the $E_2^\mathrm{high}$ mode can be interpreted as the 
phonon softening \textit{via} expansion of hexagonal $ab$-dimensions.\cite{Kim2} 
On the other hand, the broadening of the $E_2^\mathrm{high}$ mode can 
be understood in a similar context as for the XRD line-broadening and 
the near-edge absorption behavior. 
The crystal defects provide common mechanisms for the phonon lifetime ($\tauup$) shortening, and are closely related to the Raman linewidth. 
The energy-time uncertainty relation, $\Gamma/\hbar=1/\tauup$, 
indicates an inverse linear relationship between $\Gamma$ and $\tauup$. 
The wider $\Gamma$ can be attributed to the shorter $\tauup$ and in turn 
to the higher defect concentraton in the system. 
Therefore the lineshape of Raman $E_2^\mathrm{high}$ mode evidences again 
the increase of crystal defects upon the Mg-substitution in ZnO.

\section{CONCLUSIONS}

The average and local crystal structural aspects of \ZMG\/ were studied by XRD 
methods using polycrystalline samples. The composition-dependent evolutions of 
$c/a$ ratio and $u$ parameters are indicative of the polarization gradient 
along the \ZMG\/ solid solutions. It is inferred from the XRD line-broadening analysis, that the Mg atoms are homogeneously distributed over the 
wurtzite \ZMG\/ lattice without any ordering behavior. 
Raman spectroscopy provides additional evidences of $a$-lattice expansion upon 
the Mg-substitution and of the random Mg/Zn distributions. 
Although the \ZMG\/ solid solutions are obtained by an isovalent substitution, 
the band gap varies remarkably with the Mg content. The absorption edge slopes 
indicate that Mg acts as defect center near the band-edge.

\begin{acknowledgments}
Authors thank support from the U.S. National Science Foundation through 
the MRSEC program (DMR05-20415). Work at Argonne National Laboratory 
and the Advanced Photon Source was supported by the U.S. Department of Energy, 
Office of Science, Office of Basic Energy Sciences, under Contract No. DE-AC02-06CH11357. 
Authors thank Katharine Page for the synchrotron data collection, and David Clarke 
for the help in Raman measurement.
\end{acknowledgments}

\end{document}